\documentclass[
    ,final            
  ]
  {aipproc}

\layoutstyle{8x11double}

\begin{document}

\title{A Reconsideration of the $b \to s \gamma$ Decay in the Minimal Flavor Violating MSSM\footnote{Talk given by M.W. at SUSY08, Seoul, Korea}}



\author{Michael Wick}{
address={
Physik Department, 
Technische Universit\"at M\"unchen,
D-85748 Garching, Germany}}

\author{Wolfgang Altmannshofer}{
address={
Physik Department, 
Technische Universit\"at M\"unchen,
D-85748 Garching, Germany}}


\begin{abstract}
We present a MSSM study of the $b \to s \gamma$ decay in a Minimal Flavor Violating (MFV) framework, where the form of the soft SUSY breaking terms is determined by the Standard Model Yukawa couplings. In particular, we address the role of gluino contributions, which are set to zero in most studies of the MFV MSSM.

Gluino contributions can play an important role in the MFV MSSM whenever $\mu \times \tan\beta$ is large. In fact, similarly to chargino contributions, gluino contributions are $\tan\beta$ enhanced and can easily dominate charged Higgs contributions for large values of $\tan\beta$. Even though each of the separate contributions to $b \to s \gamma$ can be sizeable by itself, surprisingly no absolute lower bound can be placed on any of the relevant SUSY masses, since patterns of partial cancellations among the three competing contributions (Higgs, chargino and gluino) can occur throughout the MSSM parameter space.
\end{abstract}


\maketitle


\section{Introduction} \label{sec:intro}

The $b \to s \gamma$ decay (see \cite{Haisch} for a recent review) is a Flavor Changing Neutral Current (FCNC) process and therefore highly sensitive to New Physics (NP) effects. In fact, the agreement between the Standard Model (SM) NNLO determination of the branching ratio \cite{Misiak}
\begin{equation}\label{eq:BR_SM}
{\rm BR}(B \to X_s \gamma)^{\rm SM}_{E_\gamma > 1.6 {\rm GeV}}=(3.15 \pm 0.23) \times 10^{-4}~
\end{equation}
and the corresponding experimental average \cite{HFAG}
\begin{equation}\label{eq:BR_exp}
{\rm BR}(B \to X_s \gamma)^{\rm exp}_{E_\gamma > 1.6 {\rm GeV}}=(3.52 \pm 0.25) \times 10^{-4}~
\end{equation}
typically leads to tight constraints on the parameter space of NP models. 

One example is that of the Minimal Supersymmetric Standard Model (MSSM) with generic soft terms. In this case, flavor off-diagonal entries in the down-squark mass matrix can lead to huge NP contributions to FCNC processes like $b \to s \gamma$ and are thus constrained to unnaturally small and fine-tuned values \cite{deltas}.

A popular and elegant approach to address this ``SUSY flavor problem'' is the principle of Minimal Flavor Violation (MFV) \cite{MFV,CMFV,DAGIS}. According to the general definition~\cite{DAGIS}, the MFV hypothesis amounts to the assumption that the SM Yukawa couplings be, also in extensions of the SM, the only structures responsible for flavor (and CP) violation. In models with MFV, NP contributions to FCNC processes are thus suppressed by the same CKM factors as in the SM and therefore expected to be naturally small.

In the context of the MSSM, MFV can be realized through the following, most general ansatz for the soft SUSY breaking sector at the electro-weak scale in the SuperCKM basis \cite{DAGIS} 
\begin{eqnarray}\label{eq:softMFV}
m_Q^2 &=& \tilde m^2_Q \left( 1 + b_1 V^{\dagger} \hat Y_u^2 V + b_2 \hat Y_d^2 \right. \nonumber \\
      & & \left. + b_3 \left[ \hat Y_d^2 V^{\dagger} \hat Y_u^2 V + V^{\dagger} \hat Y_u^2 V \hat Y_d^2 \right]\right)~, \nonumber \\
m_U^2 &=& \tilde m^2_U \left( 1 + b_4 \hat Y_u^2\right)~, \nonumber \\
m_D^2 &=& \tilde m^2_D \left( 1 + b_5 \hat Y_d^2\right)~, \nonumber \\
A_u &=& \tilde A_u \left( 1 + b_6 V^* \hat Y_d^2 V^{\rm T}\right) \hat Y_u~, \nonumber \\
A_d &=& \tilde A_d \left( 1 + b_7 V^{\rm T} \hat Y_u^2 V^*\right) \hat Y_d~.
\end{eqnarray}
In the above expressions, $\tilde m^2_Q$, $\tilde m^2_U$, $\tilde m^2_D$ and $\tilde A_u$, $\tilde A_d$ represent overall mass scales for the squark bilinear and trilinear terms respectively, the coefficients $b_i$ are proportionality factors of order 1, $\hat Y_{u,d} = {\rm diag}(\lambda_{u,d},\lambda_{c,s},\lambda_{t,b})$ are the diagonal Yukawa matrices and $V$ is the CKM matrix. The coefficients $b_1$, $b_3$ and $b_7$ lead to flavor non-diagonal terms in the down squark mass matrix, which in turn imply the existence of gluino contributions to FCNC processes like $b \to s \gamma$ that, however, are proportional to the respective CKM matrix elements.

In previous phenomenological studies of the MFV MSSM adopting the general definition~\cite{DAGIS}, it was found that NP contributions to many observables are small \cite{IMPST,ABG,ABGW}.
Sizeable NP effects can however still arise in helicity suppressed decays like $B_s \to \mu^+ \mu^-$, $B^+ \to \tau^+ \bar\nu$ and also in $b \to s \gamma$.
A reconsideration of the $b \to s \gamma$ decay in the MFV MSSM, and in particular its constraining power on the MSSM parameter space, is the general aim of our work.

\section{Contributions to $b \to s \gamma$} \label{sec:contributions}

Within a MFV framework, the process $b \to s \gamma$ can be described by the $\Delta F = 1$ effective Hamiltonian
\begin{equation} \label{eq:Heff}
H_{\rm eff} = - \frac{4 G_F}{\sqrt 2} V_{ts}^* V_{tb} \sum_{i=1}^8 C_i(\mu) Q_i~,
\end{equation}
with the operators $Q_i$ given, e.g., in \cite{CMM}. In the subsequent discussion we concentrate on the NP contributions from the magnetic and chromomagnetic operators 
\begin{eqnarray} \label{eq:O78}
Q_7 &=& \frac{e}{16 \pi^2} m_b (\bar s \sigma^{\mu \nu} P_{R} b) F_{\mu \nu}~, \nonumber \\
Q_8 &=& \frac{g_s}{16 \pi^2} m_b (\bar s \sigma^{\mu \nu} T^A P_{R} b) G_{\mu \nu}^A~,
\end{eqnarray}
which are by far the dominant ones. The one-loop SUSY contributions to the corresponding Wilson coefficients can be decomposed in the following way
\begin{equation} \label{eq:C78}
C^{\rm NP}_{7,8} = C_{7,8}^{H^\pm} + C_{7,8}^{\tilde \chi^\pm} + C_{7,8}^{\tilde g} + C_{7,8}^{\tilde \chi^0}~,
\end{equation}
where the various contributions on the r.h.s. arise from penguin diagrams with charged Higgs-up type quarks, chargino-up type squarks, gluino-down type squarks and neutralino-down type squarks, respectively.
These contributions have been extensively studied in the literature for various SUSY frameworks both at LO~\cite{LO} and partly also at NLO~\cite{NLO,DAGIS}.

In the remainder of this section we qualitatively discuss the Higgs, the chargino and, in particular, the gluino contributions to the $b \to s \gamma$ decay in the context of the MFV MSSM at moderate $\tan\beta$. Neutralino penguins always give a completely negligible contribution and they will be given no further consideration.

To transparently display the dependencies of the various contributions on the SUSY parameters we use the Mass Insertion Approximation (MIA). In our numerical analysis instead, all mass matrices are diagonalized exactly.

\paragraph{Higgs contributions} \label{sec:higgs}

The contributions from charged Higgs bosons always interfere constructively with the SM ones, thus increasing the branching ratio with respect to the SM prediction. At leading order, one finds\footnote{Here and in what follows, we neglect terms suppressed by $\cot^2\beta$.}
\begin{equation} \label{eq:Higgs}
C_{7,8}^{H^\pm} \simeq f_{7,8}(x_t)~,
\end{equation}
where $x_t = m_t^2 / M_{H^\pm}^2$, with $M_{H^\pm}$ the charged Higgs mass and $f_7(1)=-7/36$, $f_8(1)=-1/6$.

\paragraph{Chargino contributions} \label{sec:chargino}

\begin{figure}[t]
\includegraphics[scale=0.5]{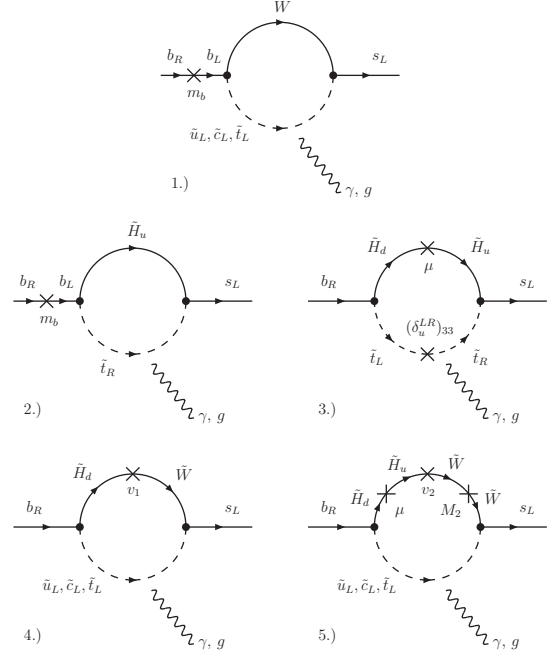}
\caption{Chargino contributions to $C_{7,8}$ in the MIA. The $(\delta_u^{LR})_{33}$ in diagram $3)$ symbolizes a left-right mass insertion in the up squark mass matrix. 
}
\label{fig:cha-MIA}
\end{figure}
In the MIA, we identify five chargino diagrams, (see Fig.~\ref{fig:cha-MIA}) that contribute to $b \to s \gamma$. The corresponding approximate expression for the Wilson coefficients read\footnote{In obtaining these expressions, all but the top Yukawa coupling were neglected in the soft terms~(\ref{eq:softMFV}). In case of large $\tan\beta$ also the bottom Yukawa would have to be kept. In addition, $\tan\beta$ enhanced corrections to the Yukawa couplings and the CKM matrix would have to be included.}
\begin{eqnarray} \label{eq:chargino}
C_{7,8}^{\tilde \chi^\pm} &\!\!\! \simeq &\!\!\! (b_1 \lambda_t^2) (M_W^2 / \tilde m_Q^2) g^1_{7,8}(x^2_Q) + (m_t^2 / \tilde m_U^2) g^2_{7,8}(x^\mu_R) \nonumber \\
&\!\!\! &\!\!\! +~(m_t^2 / \bar m_{\tilde u}^4) \mu \left( \mu + \tilde A_u \tan\beta \right) g^3_{7,8}(x^\mu) \nonumber \\
&\!\!\! &\!\!\! +~(b_1 \lambda_t^2) (M_W^2 / \tilde m_Q^4) \tan\beta \mu M_2 g^4_{7,8}(x^2_Q, x^\mu_Q) \nonumber \\
&\!\!\! &\!\!\! +~(b_1 \lambda_t^2) (M_W^2 / \tilde m_Q^2) g^5_{7,8}(x^2_Q, x^\mu_Q)~,
\end{eqnarray}
with $g^{1\dots 5}_{7,8}$ loop functions depending on the mass ratios 
\begin{equation}
x^2_Q = \frac{M_2^2}{\tilde m_Q^2} ~,~~ x^\mu_Q = \frac{\mu^2}{\tilde m_Q^2} ~,~~ x^\mu_R = \frac{\mu^2}{m^2_{\tilde t_R}} ~,~~ x^\mu = \frac{\mu^2}{\bar m^2_{\tilde u}}
\end{equation}
and $\bar m^2_{\tilde u} = m_{\tilde t_L} m_{\tilde t_R}$, where $m_{\tilde t_{L(R)}}$ is the left(right) handed top squark mass. For all masses degenerate, one has $g^1_7(1)=-1/30$, $g^2_7(1)=5/144$, $g^3_7(1)=5/72$, $g^4_7(1,1)=-13/90$, $g^5_7(1,1)=-1/180$, $g^1_8(1)=-1/40$, $g^2_8(1)=1/48$, $g^3_8(1)=1/24$, $g^4_8(1,1)=-1/15$ and $g^5_8(1,1)=1/60$.

Diagrams 1), 4) and 5) of Fig.~\ref{fig:cha-MIA} introduce a dependence on the MFV parameter $b_1$. In fact, $b_1$ does not provide any flavor violating structures in the up squark masses, but leads to a splitting between the left handed stop mass $m^2_{\tilde t_L} \approx \tilde m^2_Q(1 + b_1 \lambda_t^2)$ and the left handed up and charm squark masses $m^2_{\tilde u_L} \approx m^2_{\tilde c_L} \approx \tilde m^2_Q$. Without this splitting, such contributions would be highly suppressed by the super GIM mechanism.

\paragraph{Gluino contributions} \label{sec:gluino}

\begin{figure}[t]
\includegraphics[scale=0.5]{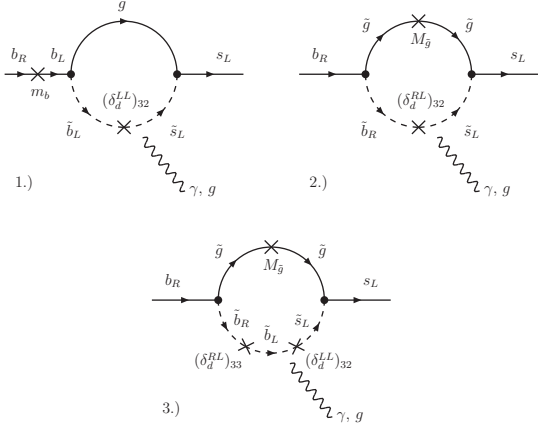}
\caption{Gluino contributions to $C_{7,8}$ in the MIA. The various $(\delta_d)$ symbolize mass insertions in the down squark mass matrix, that are determined by the MFV structure of the soft terms~(\ref{eq:softMFV}).
}
\label{fig:glu-MIA}
\end{figure}
In contrast to the up squark mass matrix, the down squark mass matrix contains flavor off-diagonal entries. These off-diagonal entries are required for the gluino contributions to $b \to s \gamma$. In the MIA there are three possible diagrams, reported in Fig.~\ref{fig:glu-MIA}, that lead to the following expressions\footnote{As for the chargino contributions, these expressions are strictly speaking only valid for moderate values of $\tan\beta$.}
\begin{eqnarray} \label{eq:gluino}
C_{7,8}^{\tilde g} &\!\!\! \simeq &\!\!\! (g_3^2/g_2^2) \left[ (b_1 \lambda_t^2)(M_W^2/\tilde m_Q^2) h^1_{7,8}(x^g_Q) \right. \nonumber \\
&\!\!\! &\!\!\! +~ (b_7 \lambda_t^2)(M_W^2 / \bar m_{\tilde d}^4) \tilde A_d M_{\tilde g} h^2_{7,8}(x^g) \nonumber \\
&\!\!\! &\!\!\! +~ (b_1 \lambda_t^2)(M_W^2 /  \bar m_{\tilde d}^4) \left( \mu \tan\beta + \tilde A_d (1 + b_7 \lambda_t^2) \right) \nonumber \\
&\!\!\! &\!\!\! \left. ~~~~\times~ M_{\tilde g} h^3_{7,8}(x^g) \right]~,
\end{eqnarray}
with $h^{1\dots 3}_{7,8}$ loop functions depending on the mass ratios
\begin{equation}
x^g_Q = \frac{M_{\tilde g}^2}{\tilde m_Q^2} ~,~~ x^g = \frac{M_{\tilde g}^2}{\bar m_{\tilde d}^2}~,
\end{equation}
with $\bar m_{\tilde d}^2 = \tilde m_Q \tilde m_D$ and $h^1_7(1)=1/45$, $h^2_7(1)=-2/27$, $h^3_7(1)=2/45$, $h^1_8(1)=7/120$, $h^2_8(1)=-5/18$ and $h^3_8(1)=7/60$.

Diagram 3) of Fig.~\ref{fig:glu-MIA} contains both a flavor violating, chirality conserving left-left mass insertion and a flavor conserving right-left mixing mass insertion that introduces a term proportional to the product $\mu \times \tan\beta$. In fact, the above expressions for $C_{7,8}^{\tilde g}$ do not decouple with $\mu$ and therefore, whenever $\mu \times \tan\beta$ is large, gluino contributions can give sizeable effects in the $b \to s \gamma$ amplitude. Similar conclusions have also been drawn in~\cite{DK}.

\section{Numerical Analysis} \label{sec:analysis}

\begin{figure}[t]
\includegraphics[scale=0.5]{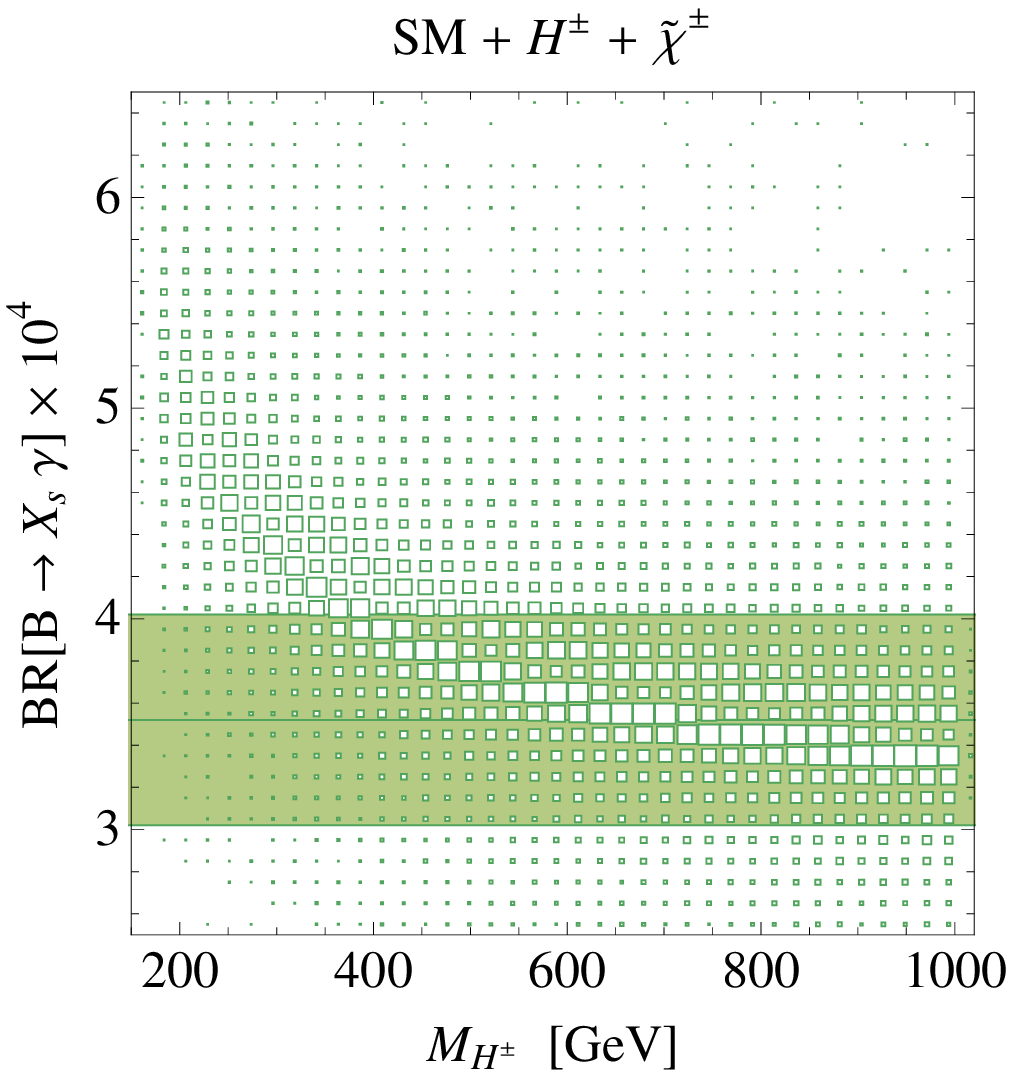}
\includegraphics[scale=0.5]{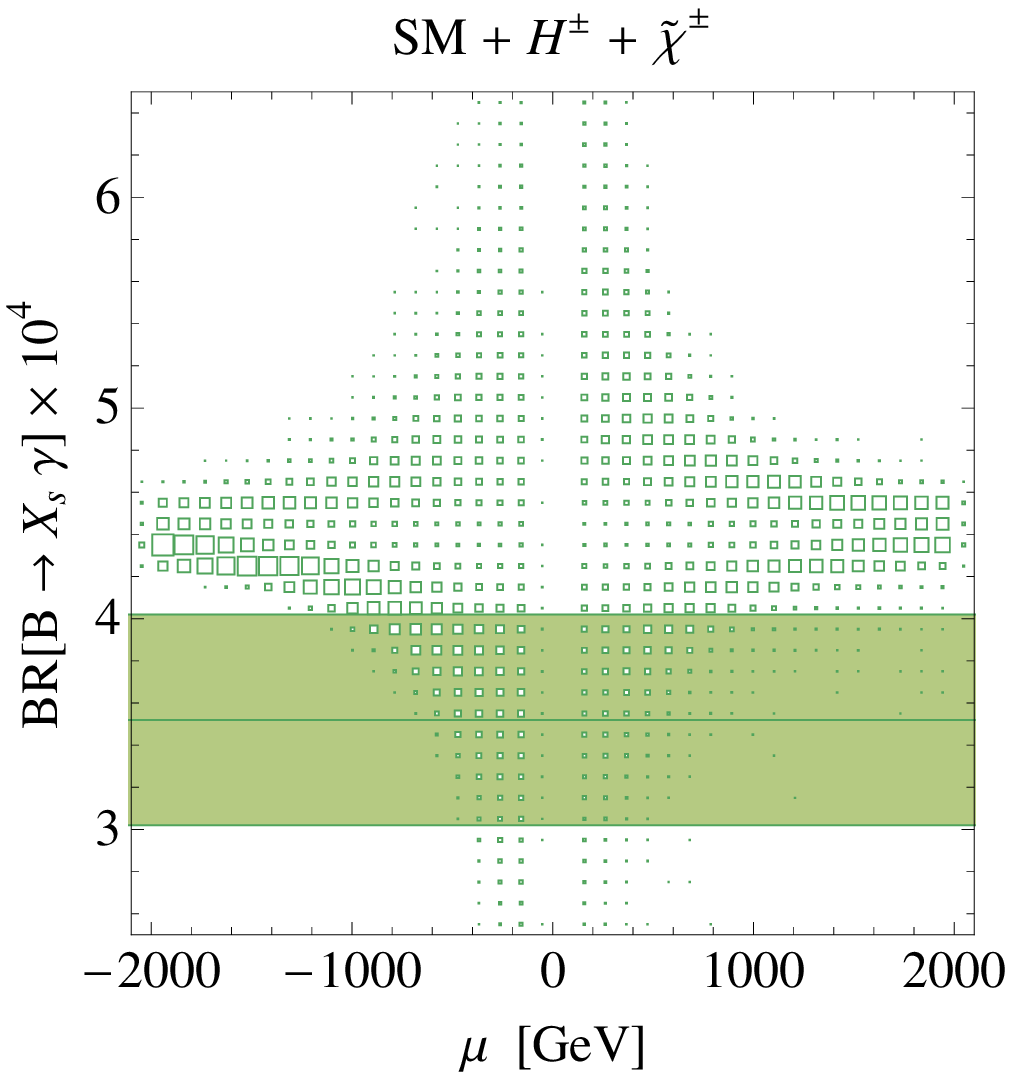}
\includegraphics[scale=0.5]{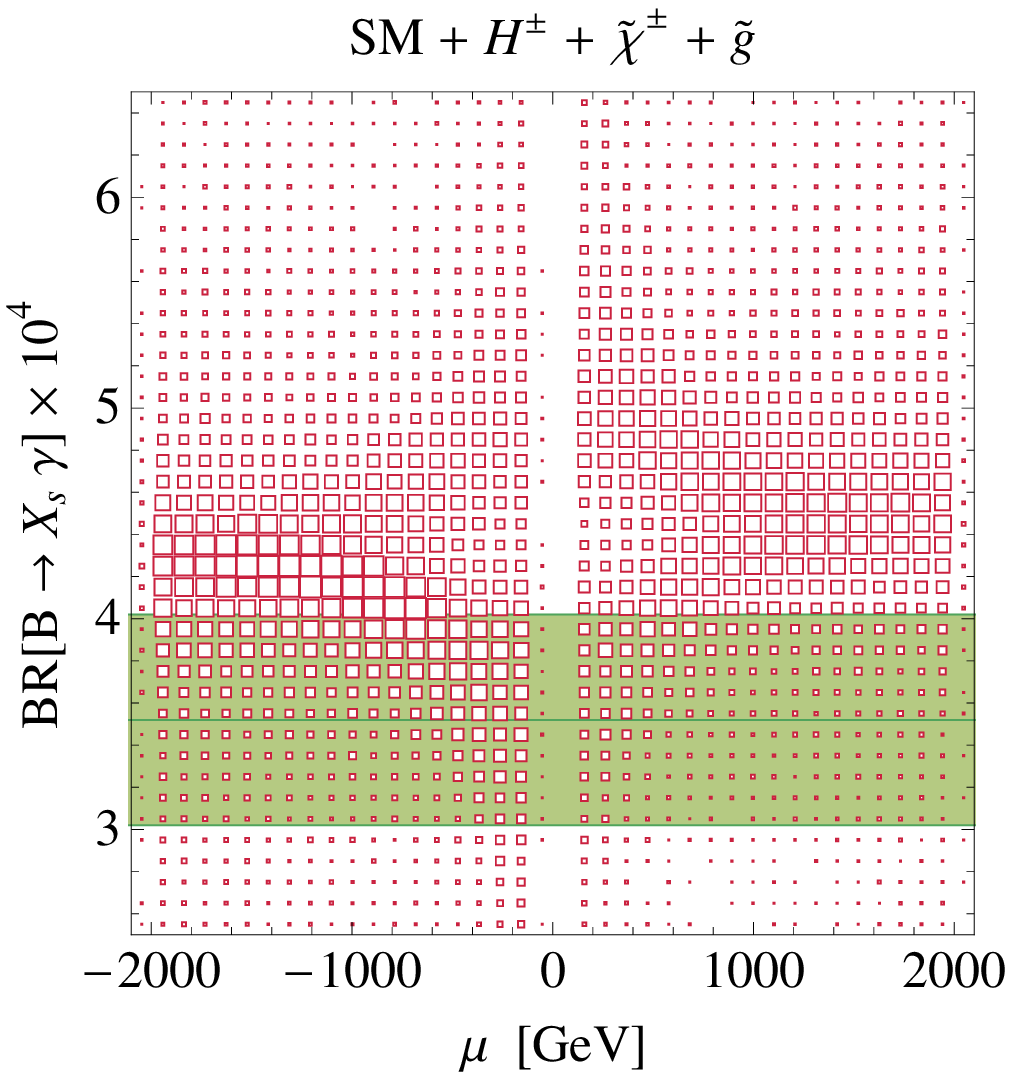}
\caption{BR$(B \to X_s \gamma)$ vs. $M_{H^\pm}$ and $\mu$. The horizontal bands indicate the experimental $2\sigma$ range. In the plots on the left and in the middle, SM, Higgs and chargino contributions are included, while the plot on the right also includes gluino contributions.}
\label{fig:plots}
\end{figure}
Starting from the Wilson coefficients discussed in the previous section, we calculate the branching ratio BR$(B \to X_s \gamma)$ using \verb|SusyBSG|~\cite{SusyBSG}. In the numerical scan we choose the following ranges for the relevant SUSY parameters
\begin{eqnarray}
M_2, M_{\tilde g}, M_{H^\pm} < 1 {\rm TeV} &;& -2  {\rm TeV} < \mu < 2 {\rm TeV} \nonumber \\
\tilde m_Q, \tilde m_U, \tilde m_D < 1.5 {\rm TeV} &;& -3  {\rm TeV} < \tilde A_u, \tilde A_d < 3 {\rm TeV} \nonumber \\
\tan\beta = 5 &;& -1 < b_1, \dots, b_7 < 1
\end{eqnarray}
and impose constraints from direct searches for SUSY particles, the lightest Higgs boson mass and FCNC observables, like $\Delta M_s / \Delta M_d$ and BR$(B_s \to \mu^+ \mu^-)$.

In the left plot of Fig.~\ref{fig:plots}, it is shown BR$(B \to X_s \gamma)$ vs. $M_{H^\pm}$, not yet including gluino contributions. As it is well known, no absolute bound on $M_{H^\pm}$ can be put in the MSSM, as chargino contributions can cancel those from Higgses. For the plot in the middle we then fix $M_{H^\pm} = 300$GeV and show BR$(B \to X_s \gamma)$ vs. $\mu$, still not including gluino contributions. In that case, one finds that for large values of $\mu$, chargino contributions are not large enough anymore to cancel the Higgs contributions and thus, large values for $\mu$ seem disfavored if the charged Higgs is light.
However, once gluino contributions are included, also for large $\mu$ cancellations are again possible (see right plot of Fig.~\ref{fig:plots}).

In fact, we find that cancellations among Higgs, chargino and gluino contributions can occur throughout the parameter space and no absolute lower bound can be placed on any of the relevant SUSY masses.

\section{Conclusions} \label{sec:conclusinos}

The general MFV ansatz for the soft SUSY breaking terms leads to gluino contributions to FCNCs. These gluino contributions can be sizeable in $b \to s \gamma$ for large values of $\mu \times \tan\beta$, due to a large flavor conserving right-left mass insertion, followed by a flavor changing left-left insertion. As it then turns out, the different contributions in the MFV MSSM - namely Higgs, chargino and gluino contributions - are naturally competitive and cancellation patterns among them are not difficult to achieve. 
In particular, we find that although the $b \to s \gamma$ decay acts as a severe constraint, it is nonetheless impossible to obtain model-independent bounds on SUSY masses from this decay alone.


\begin{theacknowledgments}
We thank Diego Guadagnoli and Andrzej Buras for useful discussions and comments on the manuscript.
This work was supported by the German Bundesministerium f\"ur Bildung und Forschung under contract 05HT6WOA.
We also acknowledge support by the Graduiertenkolleg GRK 1054.
\end{theacknowledgments}







\end{document}